\documentclass[preprint,12pt,3p]{elsarticle}
\usepackage{epsfig,amssymb,amsfonts,amsmath,mathtools,bm,color}
\biboptions{sort&compress}

\usepackage{lineno,hyperref,slashed}
\modulolinenumbers[5]

\usepackage{graphicx}
\usepackage{orcidlink}

\usepackage{bm}
\usepackage{braket}
\usepackage{multirow}
\usepackage{dsfont}
\usepackage{mathtools}
\usepackage{xcolor}
\usepackage[normalem]{ulem}

\renewcommand{\vec}[1]{\mbox{\boldmath$#1$\unboldmath}}
\newcommand{\Tch}{T_{\rm ch}}

\newcommand{\vp}{\varphi}

\newcommand{\be}{\begin{equation}}
\newcommand{\ee}{\end{equation}}
\newcommand{\bea}{\begin{eqnarray}}
\newcommand{\eea}{\end{eqnarray}}
\newcommand{\beas}{\begin{eqnarray*}}
\newcommand{\eeas}{\end{eqnarray*}}
\newcommand{\ds}{\displaystyle}
\newcommand{\vep}{{\bm p}}
\newcommand{\vek}{{\bm k}}

\newcommand{\vex}{{\bm x}}
\newcommand{\vey}{{\bm y}}

\newcommand{\veA}{{\bm A}}

\newcommand{\muIR}{\mu_{\rm IR}}

\def\vec#1{\boldsymbol{#1}}

\begin{document}

\begin{frontmatter}

\title{Chiral symmetry restoration at finite temperature in a model with manifest confinement}

\author[1]{L. Ya. Glozman}
\author[2,3]{A. V. Nefediev\orcidlink{0000-0002-9988-9430}}
\author[1]{R. Wagenbrunn}

\address[1]{Institute of Physics, University of Graz, A-8010 Graz, Austria,
e-mails: leonid.glozman@uni-graz.at; robert@wagenbrunn.net}

\address[2]{Jo{\v z}ef Stefan Institute, Jamova 39, 1000, Ljubljana, Slovenia}

\address[3]{CeFEMA, Center of Physics and Engineering of Advanced Materials, Instituto Superior T{\'e}cnico, Av. Rovisco Pais, 1 1049-001 Lisboa, Portugal,
e-mail: a.nefediev@gmail.com}

\begin{abstract}
Multiple lattice evidences support the existence of a
confining but chirally symmetric regime of QCD above the
chiral symmetry restoration crossover at $\Tch \simeq 155$ MeV.
This regime is characterised by an approximate chiral spin
symmetry of the partition function, which is a symmetry of the 
colour charge and the confining electric part of the QCD Lagrangian.
It is traditionally believed that confinement should automatically induce
spontaneous breaking of chiral symmetry, which would preclude the
existence of a confining but chirally symmetric regime of QCD at high
temperatures. We employ a well-known solvable quark model for QCD in 3+1
dimensions that is chirally symmetric and manifestly confining and
argue that while confinement indeed induces dynamical breaking
of chiral symmetry at $T=0$, a chiral restoration phase transition takes place
at some critical temperature $\Tch$. Above this temperature, the spectrum
of the model consists of chirally symmetric hadrons with approximate
chiral spin symmetry.
\end{abstract}


\end{frontmatter}


\section{Introduction}

It had been widely believed for a long time that, in QCD upon heating,
the pseudo-critical temperature $\Tch \simeq 155$ MeV \cite{Aoki:2009sc,HotQCD:2018pds,Borsanyi:2020fev} should be a common temperature for the chiral restoration and deconfinement. Meanwhile, lattice studies at $T=0$ performed on the field configurations with artificially truncated near-zero modes of the Dirac operator \cite{Denissenya:2014poa,Denissenya:2014ywa} suggested approximate emergent symmetries of the spectrum of hadrons \cite{Glozman:2014mka,Glozman:2015qva} associated with the symmetries of the colour charge and confining
electric part of the QCD Lagrangian --- $SU(2)_{CS}$ chiral spin symmetry and its flavour extension $SU(2N_F)$. For a review on these symmetries and their implications in hot QCD see Ref.~\cite{Glozman:2022zpy}.

Further lattice studies indicated the emergence of approximate chiral spin $SU(2)_{CS}$ and $SU(4)$ symmetries also above $\Tch$ thus suggesting that the theory should still be in the confining regime \cite{Rohrhofer:2017grg,Rohrhofer:2019qwq,Rohrhofer:2019qal,Chiu:2023hnm}. In other words, in QCD upon heating, the
quark-gluon plasma regime does not immediately follow the hadron gas but there exists an intermediate confining regime, called stringy fluid, with restored chiral symmetry and approximate chiral spin symmetry. Also, there are additional not related to symmetry lattice evidences supporting the existence of this intermediate regime \cite{Glozman:2022lda,Lowdon:2022xcl,Bala:2023iqu,Cohen:2024ffx}.

The centre symmetry of the gauge action is explicitly broken by quark loops but becomes exact in the large-$N_c$ limit. Then QCD in the combined large-$N_c$ and chiral limit features two distinct symmetries, chiral symmetry and centre symmetry, which allow one to define unambiguously possible phases of the theory with confinement or deconfinement and with spontaneously broken or restored chiral symmetry. It was suggested in Ref.~\cite{Cohen:2023hbq} that the three regimes of QCD, that are connected through smooth crossovers in the real world with $N_c=3$ and a small quark mass, might become distinct phases separated by phase transitions in the large-$N_c$ world with massless quarks.

The existence of the above putative chirally symmetric but confining phase of QCD raises questions since one of the heritages of the MIT bag model is a believe that confinement necessarily induces breaking of chiral symmetry. A similar statement was made by Casher~\cite{Casher:1979vw}. However, as demonstrated in
Refs.~\cite{Glozman:2007tv,Glozman:2009sa}, this inference does not hold
in dense hadronic systems. We also note that the 't~Hooft anomaly matching conditions \cite{tHooft:1979rat} and the large-$N_c$ Coleman-Witten theorem for chiral symmetry breaking \cite{Coleman:1980mx} do not relate the notion
of confinement to the centre symmetry. Importantly, they both exploit Lorentz invariance and, therefore, are by construction restricted to zero temperature only. In addition, they rely on a ``naive'' definition of confinement as an asymptotic existence of only colour-singlet hadronic states. This definition does not apply to dense hadronic systems.

In this paper we employ a solvable chirally symmetric quark model with a manifestly confining interaction potential, that necessarily induces spontaneous breaking of chiral symmetry at zero temperature, and demonstrate that the aforementioned confining but chirally symmetric phase of the theory can exist at high temperatures.

\section{Chiral restoration at finite temperature in Nambu--Jona-Lasinio model}

We start from a brief reminder of the mechanism of chiral symmetry
restoration at a finite temperature in the Nambu--Jona-Lasinio model described by the interaction Hamiltonian density \cite{Vogl:1991qt,Klevansky:1992qe,Hatsuda:1994pi}
\be
H=-G\left[(\bar \psi \psi)^2+(\bar \psi i\gamma_5\vec{\tau}\psi)^2\right].
\label{NJL}
\ee
If the strength of the effective interaction $G$ exceeds some critical level, a nonlinear gap equation,
\be
M=-2G\braket{\bar \psi\psi},\quad
\braket{\bar \psi\psi}=-\frac{N_c}{\pi^2}
\int_0^{\Lambda_\chi} {p}^2 d{p} \frac{M}{\sqrt{{p}^2+ M^2}},
\label{gap2}
\ee
admits a nontrivial solution with $ \braket{\bar \psi\psi}\neq 0$, which implies that the initial vacuum is rearranged and chiral symmetry in the ``physical'' vacuum is dynamically broken. Thus, appearance of a nonvanishing quark condensate results in a gap in the spectrum of elementary excitations. Dressed quarks, regarded as quasiparticles in the Bogoliubov sense (that is, as coherent superpositions of the bare particle-antiparticle excitations), acquire a dynamically generated mass $M$ while the new vacuum is again trivial in terms of these quasiparticles and can be annihilated by the corresponding quark and antiquark annihilation operators.

Consider now chiral symmetry restoration in the vacuum at high temperatures. Equation (\ref{gap2}) describes the chiral condensate at $T=0$. A finite temperature $T$ induces particle-antiparticle excitations according to the standard Fermi-Dirac distribution functions for the quarks and antiquarks (that coincide at a vanishing chemical potential),
\be
n_p=\bar{n}_p=\left(1 + e^{\sqrt{p^2 + M^2}/T}\right)^{-1}\mathop{\to}_{T\to\infty}\frac12,
\label{FD}
\ee
this way affecting the quasiparticle gas in the vacuum, so that now
\be
\braket{\bar \psi\psi}=-\frac{N_c}{\pi^2}
\int_0^{\Lambda_\chi} {p}^2 dp\frac{M}{\sqrt{p^2 + M^2}}
[1-n_p-\bar{n}_p].
\label{gap3}
\ee

At some critical temperature $\Tch$ the mass-gap equation (\ref{gap3})
ceases to have a nontrivial solution and chiral symmetry gets restored in the vacuum. The physical picture behind this chiral symmetry restoration is related to Pauli blocking of the levels, necessary for the formation of the condensate, by the thermal excitations of quarks and antiquarks. Meanwhile, the formal (mathematical) reason is that the function $1-2n_p$, with the distribution function $n_p$ in Eq.~(\ref{FD}), effectively damps the integral on the right-hand side of the gap equation (\ref{gap3}).

The outlined mechanism of chiral symmetry restoration at finite temperatures is precisely the same as the mechanism of the second order
phase transition in the Bardeen-Cooper-Schriefer theory
of superconductivity \cite{Bardeen:1957kj,Bardeen:1957mv}
. Below we demonstrate that a similar mechanism is also operative in a manifestly confining chirally symmetric model for QCD.

\section{Confining chirally symmetric model in 3+1 dimensions}

\subsection{Symmetries of the QCD Hamiltonian in the Coulomb gauge}

Consider the Minkowskian QCD Hamiltonian in the Coulomb gauge in the rest frame system in the chiral limit \cite{Christ:1980ku},
\be
H_{\rm QCD}=H_E+H_B+\int d^3 x\;\psi^\dag(\vex)(-i{\bm\alpha}\cdot{\bm\nabla})\psi(\vex)+H_T+H_C,
\label{ham}
\ee
with the transverse (magnetic) and instantaneous ``Coulombic'' interactions,
\be
H_T=-g\int d^3 x\;\psi^\dag(\vex){\bm\alpha}\cdot\veA^a(\vex)t^a \psi(\vex),
\ee
and
\be
H_C=\frac{g^2}2\int d^3\;x d^3 y\;J^{-1}\rho^a(\vex)F^{ab}(\vex,\vey) J\rho^b(\vey),
\label{coul}
\ee
respectively. Here $J$ is the Faddeev-Popov determinant, $\rho^a(\vex)$ is the colour-charge density of quarks and gluons at the space points $\vex$, and $F^{ab}(\vex,\vey)$ is the ``Coulombic'' kernel.

The kinetic and transverse parts of the Hamiltonian (\ref{ham}) are
chirally symmetric while the symmetry of the confining ``Coulombic'' part (\ref{coul}) is $SU(2N_F) \times SU(2N_F)$. Indeed, the quark colour charge density operator is $SU(2N_F)$-symmetric and the $SU(2N_F)$ rotations performed at the spatial points $\vex$ and $\vey$ are independent \cite{Glozman:2022zpy}.
 
\subsection{Confining chiral quark model at $T=0$}

A simplified model for strong interactions --- large-$N_c$ QCD in 1+1 dimensions --- was suggested by 't~Hooft 50 years ago \cite{tHooft:1974pnl}. Because of the reduced number of spatial dimensions, in this model, there is no magnetic field and the only gluonic interaction between quarks is provided by a confining potential that rises linearly with the interquark separation. The Hamiltonian (\ref{ham}) is reduced then to a confined theory without the Faddeev-Popov determinant and the transverse part \cite{Bars:1977ud}. An incontestable success of the 't~Hooft model in providing insight into various features inherent to full QCD in 3+1 dimensions stimulated multiple attempts to study the latter using confining chiral quark models based on the same truncations as those employed in the 't~Hooft model. In particular, considering the model in the spirit of large-$N_c$ QCD one can disregard the string breaking mechanism. In addition, ignoring for simplicity the magnetic part of the interaction, one arrives at a 3+1-dimensional model reminiscent of the two-dimensional 't~Hooft model,
\be
H=\int d^3x\;\bar{\psi}(\vex,t)\left(-i\vec{\gamma}\cdot
{\bm\nabla}+m\right)\psi(\vex,t)
+\frac{1}{2} \int d^3x\;d^3y\;\rho^a(\vex)F^{ab}(\vex,\vey)\rho^b(\vey),
\label{HGNJL}
\ee
with the colour charge density $\rho^a(\vex)=\psi^\dag(\vex)\frac{\lambda^a}{2}\psi(\vex)$ and an instantaneous confining kernel $F^{ab}(\vex,\vey)=\delta^{ab}V_0(|\vex-\vey|)$. This quark model has a rich history --- see, for example, Refs.~\cite{Amer:1983qa,LeYaouanc:1984ntu,Adler:1984ri,Bicudo:1989sh,Bicudo:2002eu,Llanes-Estrada:1999nat,Nefediev:2004by,Alkofer:2005ug,Wagenbrunn:2007ie} and references therein. Importantly, all qualitative conclusions deduced using the model (\ref{HGNJL}) are insensitive to a particular choice of the confining potential. Then, while the appearance of a linearly rising potential between quarks in 1+1 dimensions is a natural consequence of the form of the two-dimensional gluon propagator in the Coulomb gauge, the microscopic origins of linear confinement in 3+1 dimensions are more obscure (see Ref.~\cite{Nguyen:2024ikq} for recent insights and references). Still, in practical calculations, a linearly rising form
$V_0(|\vex-\vey|) \sim|\vex-\vey|$ is typically celebrated as the most phenomenologically adequate choice. Such linear confinement was also obtained as a result of variational calculations in quenched Coulomb-gauge QCD \cite{Szczepaniak:2001rg,Feuchter:2004mk}.

As reported in the pioneering paper \cite{Amer:1983qa}, the chirally invariant vacuum of the theory described by the Hamiltonian (\ref{HGNJL}) is unstable. An appropriate language to use is a Bogoliubov-Valatin transformation of the quark field,
\be
\psi(\vex,t)=\sum_{s=\uparrow,\downarrow}\int\frac{d^3p}{(2\pi)
^3}e^{i\vep\vex}\left[e^{-iE_pt}b_s(\vep)u_s(\vep)+e^{iE_pt}d_s^\dagger(-\vep)v_s(-\vep)\right],
\label{psi}
\ee
with $E_p$ for the dispersion law of the dressed fermions, parametrised in terms of a function of the 3-momentum $\vp(p)\equiv\vp_p$ known as chiral angle,
\be \left\{
\begin{array}{rcl}
u(\vep)&=&\ds\frac{1}{\sqrt{2}}\left[\sqrt{1+\sin\vp_p}+
\sqrt{1-\sin\vp_p}\;(\vec{\alpha}\hat{\vep})\right]u(0),\\
v(-\vep)&=&\ds\frac{1}{\sqrt{2}}\left[\sqrt{1+\sin\vp_p}-
\sqrt{1-\sin\vp_p}\;(\vec{\alpha}\hat{\vep})\right]v(0).
\end{array}
\right.
\label{uandv}
\ee
By convention, the chiral angle typically varies in the range
$-\frac{\pi}{2}<\vp_p\leqslant \frac{\pi}{2}$ with the boundary
conditions $\vp(0)=\frac{\pi}{2}$, $\vp(p\to\infty)\to 0$.
A particular profile of the chiral angle comes as the solution of a nonlinear integral equation derived below and known as mass-gap equation.

We start from the case of $T=0$. Then the mass-gap equation can be derived from the requirement that the vacuum energy is a minimum in terms of the dressed quark field (\ref{psi}) \cite{Amer:1983qa,LeYaouanc:1984ntu,Adler:1984ri}. Alternatively, it follows from the requirement that the term in the Hamiltonian (\ref{HGNJL}) quadratic in the dressed quark/antiquark creation and annihilation operators should not contain anomalous ($\sim b^\dagger d^\dagger$, $db$) contributions \cite{Bicudo:1989sh}. For future convenience, here we derive the mass-gap equation for the chiral angle from the Schwinger-Dyson series for the dressed quark propagator \cite{Bars:1977ud},
\be
S(p_0,\vep)=\frac{\Lambda_+(\vep )\gamma_0}{p_0-E_p+i\epsilon}+
\frac{{\Lambda_-}(\vep )\gamma_0}{p_0+E_p-i\epsilon},
\label{Feynman}
\ee
where the positive- and negative-energy projectors for the dressed fermion read
\be
\Lambda_\pm(\vep )=\frac12[1\pm\gamma_0\sin\vp_p\pm(\boldsymbol{\alpha}\hat{\vep })\cos\vp_p],
\label{Lpm}
\ee
with $\hat{\vep}$ for the unit vector in the direction of the momentum $\vep$. The propagator (\ref{Feynman}) comes as a sum of the Dyson series,
\be
S=S_0+S_0\Sigma S_0+S_0\Sigma S_0\Sigma S_0+\ldots=S_0+S_0\Sigma S,
\label{Ds}
\ee
where
\be
S_0(p_0,\vep)=\frac{1}{\gamma_0 p_0-\boldsymbol{\gamma}\vep-m+i\epsilon}
\label{S0T0}
\ee
is the bare quark propagator, and the self-energy function is given by an integral from the same dressed propagator in Eq.~(\ref{Feynman}),
\be
i\Sigma(\vep )=\int\frac{d^4k}{(2\pi)^4}V(\vep -\vek)\gamma_0 S(k_0,{\vek})\gamma_0,
\label{Sigma0}
\ee
where a precise definition of the confining propagator $V(\vep)$ is given later.

Solution of the coupled equations (\ref{Ds}) and (\ref{Sigma0}) can be found in the form
\be
S^{-1}(p_0,\vep )=\gamma_0p_0-({\boldsymbol{\gamma}}\hat{\vep })B_p-A_p
\label{ST0}
\ee
and
\be
\Sigma(\vep )=[A_p-m]+({\boldsymbol{\gamma}\hat{\vep }})[B_p-p],
\label{Sigma}
\ee
with the functions $A_p$ and $B_p$, parametrised via the chiral angle introduced in Eq.~(\ref{uandv}),
\bea
A_p&=&m+\frac12\int\frac{d^3k}{(2\pi)^3}V(\vep-\vek)\sin\vp_k,\nonumber\\[-2mm]
\label{AB}\\[-2mm]
B_p&=&p+\frac12\int \frac{d^3k}{(2\pi)^3}\;
(\hat{\vep}\hat{\vek})V(\vep-\vek)\cos\vp_k.\nonumber
\eea

Equation (\ref{Ds}) is satisfied if the functions $A_p$ and $B_p$ obey the condition
\be
A_p\cos\vp_p-B_p\sin\vp_p=0
\label{mge}
\ee
that is the sought mass-gap equation defining the profile of the chiral angle $\varphi_p$. The mass-gap equation with a linear potential was first solved in Ref.~\cite{Adler:1984ri}, and this solution was later re-confirmed in many subsequent papers. It was also generalised to other shapes of the confining interaction potential. With a found solution of the mass-gap equation the energy of the dressed fermion (dispersion law) introduced in Eq.~(\ref{psi}) is built as
\be
E_p=A_p\sin\vp_p+B_p\cos\vp_p.
\label{Ep}
\ee
In particular, for a non-interacting theory with $V(r)=0$, the chiral angle is
\be
\sin\vp^{(0)}_p=\frac{m}{\sqrt{p^2+m^2}},\quad
\cos\vp^{(0)}_p=\frac{p}{\sqrt{p^2+m^2}},
\label{vp0}
\ee
and, therefore, the quark dispersion law takes its free form
$E^{(0)}_p=\sqrt{p^2+m^2}$.

As follows from the definitions in Eq.~(\ref{AB}), the mass-gap equation (\ref{mge}) contains the Fourier transform of the confining propagator. However, the latter is typically ill-defined because of the singular behaviour of the potential in the infrared region (see Refs.~\cite{Amer:1983qa,Bicudo:2003cy} for a detailed study of power-like confining potentials).
Thus, in order to prescribe a physical meaning to all the quantities introduced above, one needs to impose a suitable infrared regularisation. There are many ways to do it, and the physical results should not depend on a particular regularisation scheme. Here, to illustrate the pattern, we consider a linearly rising potential $V_0$ and employ the prescription adopted in Refs.~\cite{Alkofer:2005ug,Wagenbrunn:2007ie,Bicudo:2010qp} that consists in introducing an infrared regulator to the potential and propagator in the momentum space,
\be
V(p)=\frac{8\pi\sigma}{(p^2+\muIR^2)^2},
\label{FV} 
\ee
where the $N_c$-dependent Casimir factor
$\frac{N_c^2-1}{2N_c}$ is absorbed into the
string tension $\sigma$. Notice that the sign of the confining propagator $V(p)$ is just opposite to the sign of the confining potential $V_0(p)$.
Then, upon the inverse transformation to the coordinate space,
\be
-V(r)=-\int \frac{d^3 p}{(2\pi)^3} V(p) e^{i\vec p \vec r}=-\frac{\sigma}{\muIR}e^{-\muIR r}\mathop{=}_{\muIR\to 0}-\frac{\sigma}{\muIR}+\sigma r+{\cal O}(\muIR),
\label{div}
\ee
one recovers the linear potential plus a constant term that diverges in the limit $\muIR\rightarrow 0$. This singular term must cancel in all observables while nonobservable quantities may depend on it. In particular, it is easy to verify that \cite{Wagenbrunn:2007ie}
\begin{align}
A_p=\frac{\sigma}{2\muIR}\sin\varphi_p+A_p^{\rm fin},\nonumber\\[-3mm]
\label{ABir}\\[-3mm]
B_p=\frac{\sigma}{2\muIR}\cos\varphi_p+B_p^{\rm fin},\nonumber
\end{align}
with $A_p^{\rm fin}$ and $B_p^{\rm fin}$ for infrared-finite contributions. Then it is easy to see that the mass-gap equation (\ref{mge}) is infrared-divergence-free and so is the chiral angle $\varphi_p$.
However, as can be readily inferred from Eq.~(\ref{ABir}), the function $E_p$ in Eq.~(\ref{Ep}) is infrared divergent,
\be
E_p=\frac{\sigma}{2\muIR}+\ldots,
\label{Epir}
\ee
with the ellipsis for infrared-finite terms. Consequently the single-quark Green's function (\ref{Feynman}) is also divergent in the infrared limit, that is, a single quark is repelled from the Hilbert space in agreement with natural expectations. Meanwhile, the observable colour-singlet meson masses are infrared-finite. The latter is because the infrared divergence of the single quark Green's function exactly cancels out with the infrared divergence of the kernel in the Bethe-Salpeter equation for a quark-antiquark bound state \cite{Wagenbrunn:2007ie}. As a matter of fact, such cancellation of the infrared divergences is exact in any colour-singlet observable while coloured objects are infrared divergent and hence are removed from the physical spectrum \cite{Glozman:2008fk}. The confining properties of the considered quark model are precisely the same as in 't~Hooft model --- see, for example, review \cite{Kalashnikova:2001df} and references therein.

The fact that the single-quark dispersion law (\ref{Ep}) is ill-defined in the infrared limit makes this quantity inappropriate for addressing the properties of the dressed fermion. Therefore, for practical purposes it proves convenient to define an infrared-finite dressed quark energy $\omega_p$ such that
\be
B_p\omega_p=pE_p.
\label{omegapdef}
\ee
In the no-interaction limit, one has the solution (\ref{vp0}) and then, trivially, $\omega_p^{(0)}=E_p^{(0)}=\sqrt{p^2+m^2}$ while, for a nonvanishing interaction,
\be
\omega_p=p\lim_{\muIR\to 0}\frac{E_p}{B_p}=
\frac{p}{\cos\varphi_p}=p\sqrt{1+\tan^2\varphi_p}=
\sqrt{p^2+M_p^2},
\ee
that comes as a natural generalisation of the corresponding relation in the no-interaction limit to a ``dynamical quark mass''
\be
M_p=p\tan\varphi_p.
\label{Mp}
\ee
Indeed, in the no-interaction limit, $M^{(0)}_p=m$ while a nonvanishing interaction provides an additional contribution to $M_p$ as encoded in Eq.~(\ref{Mp}). Since the chiral angle is infrared-finite, a well-defined infrared behaviour of the effective quark mass $M_p$ is evident.

The solution of the gap equation $\varphi_p$ and the shape of the dynamical mass $M_p$ in dependence on the infrared regulator $\muIR$ were studied in detail in Refs.~\cite{Wagenbrunn:2007ie,Bicudo:2010qp}. In particular, a smooth $\muIR\to 0$ limit was established numerically for both quantities. In addition, a complete meson spectrum was obtained in Ref.~\cite{Wagenbrunn:2007ie} that demonstrates an effective restoration of chiral symmetry at large hadron angular momenta $J$'s.
Moreover, one can easily see that the multiplets obtained in that work
for large $J$'s are precisely the multiplets of a larger chiral spin and $SU(4) \times SU(4)$ symmetries of confinement.

In the nontrivial vacuum defined by the solution of the mass-gap equation (\ref{mge}), the chiral condensate takes a nonzero value and spontaneously breaks chiral symmetry,\footnote{Hereinafter we work in the chiral limit of $m=0$. For $m\neq 0$, a divergent contribution to the chiral condensate that comes from the explicit-chiral-symmetry-breaking solution (\ref{vp0}) needs to be subtracted.}
\be
\braket{\bar{\psi}\psi}=-\frac{N_c}{\pi^2}\int^{\infty}_0 dp\;p^2\sin\vp_p\neq 0.
\label{Sigma1}
\ee

\subsection{Chiral restoration at finite $T$}

We now proceed to a finite temperature $T$ employing the real-time formalism. The thermal Green's function of a bare fermion acquires an additional contribution \cite{Asakawa:1989bq},
\be
S_0(p_0,\vep;T)=S_0(p_0,\vep;T=0)+
2\pi i(\gamma^\mu p_\mu+m)\delta(p^2-m^2)\Bigl[\theta(p_0)n_p^{(0)}+\theta(-p_0)\bar{n}^{(0)}_p\Bigr],
\label{S0T}
\ee
where $S_0(p_0,\vep;T=0)\equiv S_0(p_0,\vep)$ is quoted in Eq.~(\ref{S0T0}), $\theta$ is the Heaviside step-like function, and
\be
n_p^{(0)}=\frac{1}{e^{(\sqrt{p^2+m^2}-\mu)/T}+1},\quad \bar{n}^{(0)}_p=\frac{1}{e^{(\sqrt{p^2+m^2}+\mu)/T}+1}
\ee
are the Fermi-Dirac functions for the free quark and antiquark, respectively, with $\mu$ for the chemical potential. Generalisation of Eq.~(\ref{S0T}) to the dressed fermionic field (\ref{psi}) can be done using the definition of the thermal averages in terms of the dressed quarks,
\be
\braket{b_s^\dagger(\vep)b_{s'}(\vep')}=n_p\delta_{ss'}(2\pi)^3\delta^{(3)}(\vep-\vep'),\quad
\braket{d_s^\dagger(\vep)d_{s'}(\vep')}=\bar{n}_p\delta_{ss'}(2\pi)^3\delta^{(3)}(\vep-\vep'),
\ee
where now
\be
n_p=\frac{1}{e^{(\sqrt{p^2+M_p^2}-\mu)/T}+1},\quad \bar{n}_p=\frac{1}{e^{(\sqrt{p^2+M_p^2}+\mu)/T}+1},
\ee
with the effective mass of the dressed fermion defined in Eq.~(\ref{Mp}) above. Then we find
\be
S(p_0,\vep;T)=S(p_0,\vep;T=0)+
2\pi i\delta(p_0^2-E_p^2)2E_p\Bigl[\theta(p_0)n_p\Lambda_+(\vep)-\theta(-p_0)\bar{n}_p\Lambda_-(\vep)\Bigr]\gamma_0,
\label{ST}
\ee
with the zero-temperature dressed Green's function $S(p_0,\vep;T=0)\equiv S(p_0,\vep)$ defined in Eq.~(\ref{ST0}) and the projectors $\Lambda_\pm(\vep)$ introduced in Eq.~(\ref{Lpm}). For the non-interacting theory with $V(r)=0$, the Green's function (\ref{ST}) readily turns to $S_0(p_0,\vep;T)$ from Eq.~(\ref{S0T}). Equation (\ref{ST}) allows one to find the self-energy (\ref{Sigma0}) in the form
\be
\Sigma(\vep ;T)=[\tilde{A}_p-m]+(\boldsymbol {\gamma}\hat{\vep })[\tilde{B}_p-p]+\gamma_0\tilde{C}_p,
\label{SigmaT}
\ee
with
\begin{align}
\tilde{A}_p&=m+\frac12\int\frac{d^3k}{(2\pi)^3}(1-n_k-\bar{n}_k)V(\vep-\vek)\sin\vp_k,\nonumber\\
\tilde{B}_p&=p+\frac12\int \frac{d^3k}{(2\pi)^3}(1-n_k-\bar{n}_k)
(\hat{\vep}\hat{\vek})V(\vep-\vek)\cos\vp_k,
\label{tildeABC}\\
\tilde{C}_p&=-\frac12\int\frac{d^3k}{(2\pi)^3}(n_k-\bar{n}_k)V(\vep-\vek).\nonumber
\end{align}
In the derivation above, easily verifiable relations,
\be
n_p\Lambda_+(\vep)-\bar{n}_p\Lambda_-(\vep)=
\frac12(n_p+\bar{n}_p)[\Lambda_+(\vep)-\Lambda_-(\vep)]+\frac12(n_p-\bar{n}_p)[\Lambda_+(\vep)+\Lambda_-(\vep)]
\ee
and
\be
\Lambda_+(\vep)+\Lambda_-(\vep)=\gamma_0,\quad \Lambda_+(\vep)-\Lambda_-(\vep)=\sin\vp_p+(\boldsymbol{\gamma}\hat{\vep})\cos\vp_p,
\ee
were used. Alternatively, the expressions in Eq.~(\ref{tildeABC}) can be derived in the imaginary time framework \cite{Kocic:1985uq}.
In the absence of the chemical potential ($\mu=0$), $\bar{n}_p=n_p$ and, therefore, $\tilde{C}_p=0$. Then the thermal infrared-finite mass-gap equation (\ref{mge}) takes the form
\be
\tilde{A}_p\cos\vp_p-\tilde{B}_p\sin\vp_p=0.
\label{mgeT}
\ee

As the temperature $T$ increases, the Fermi-Dirac function $n_p$ approaches its infinite-$T$ limit $n_p(T\to\infty)=\frac12$ and the integral terms in the functions $\tilde{A}_p$ and $\tilde{B}_p$ gradually vanish. Consequently, the chiral angle takes its free limit (\ref{vp0}) that corresponds to the Wigner-Weyl mode of the theory at $m=0$. Thus, at $T=0$, one has the Nambu-Goldstone mode with chiral symmetry spontaneously broken while, at $T=\infty$, the theory is in the Wigner-Weyl mode with chiral symmetry manifest. Then a chiral restoration phase transition must take place at some finite temperature $\Tch$ to be numerically found from the thermal mass-gap equation (\ref{mgeT}). It may slightly vary with the particular shape of the employed confining potential. The spectrum of mesons above the chiral restoration phase transition (for $\vp_p=0$) obtained from the Bethe-Salpeter equation like the one formulated and solved in Ref.~\cite{Wagenbrunn:2007ie} will exhibit chiral symmetry and approximate $SU(4)\times SU(4)$ symmetry.

\begin{figure}[t!]
\centering
\includegraphics[width=0.6\textwidth]{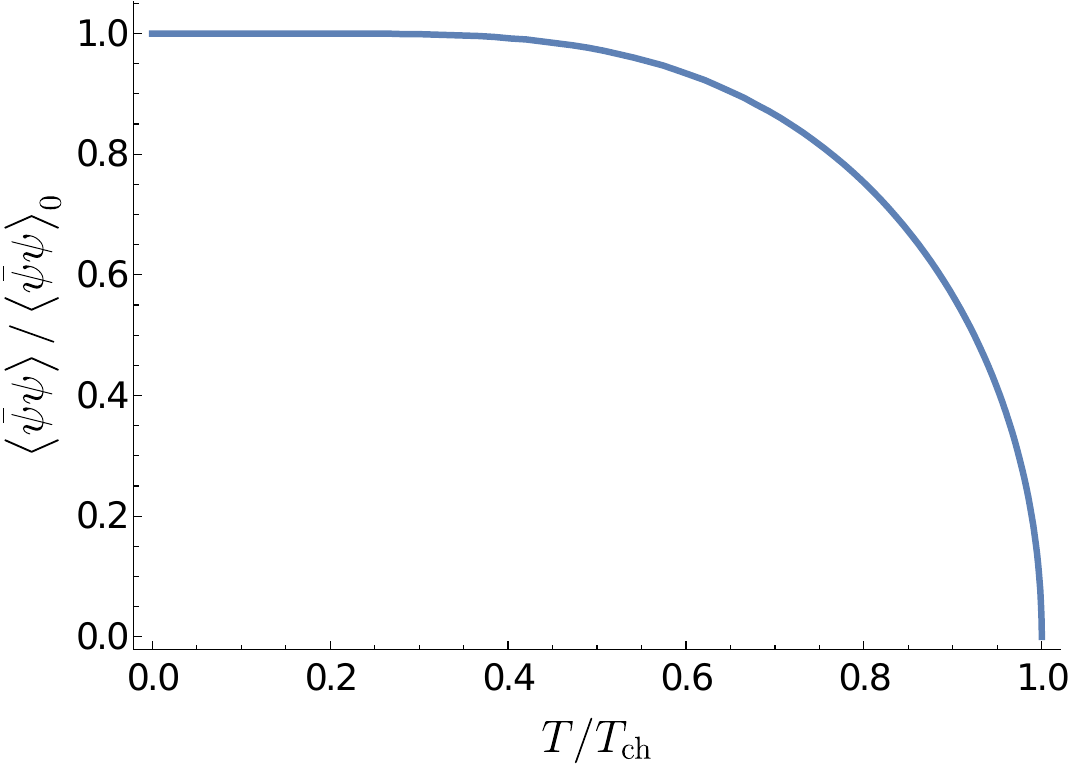}
\caption{Temperature dependence (with $\Tch$ in Eq.~\eqref{Tch}) of the chiral condensate in Eq.~\eqref{Sigma1}, normalised to that at $T=0$ in Eq.~\eqref{cond1}.}
\label{fig:cond}
\end{figure}

After the above qualitative consideration of the thermal mass-gap equation \eqref{mgeT}, we now solve this equation numerically for finite temperatures $T>0$ with linear confinement.  The result of the calculations is shown in Fig.~\ref{fig:cond} and demonstrates that, in agreement with the qualitative arguments presented above, the chiral condensate $\braket{\bar{\psi}\psi}$ takes its maximal value at $T=0$,
\be
\braket{\bar{\psi}\psi}_0\equiv \braket{\bar{\psi}\psi}_{T=0}\approx-0.0123(\sqrt{\sigma})^3,
\label{cond1}
\ee
and then decreases with the rise of the temperature until it finally vanishes at
\be
\Tch\approx 0.084\sqrt{\sigma}.
\label{Tch}
\ee
If the string tension parameter $\sigma$ is excluded from Eqs.~\eqref{cond1} and \eqref{Tch}, the relation predicted by the model reads
\be
|\braket{\bar{\psi}\psi}_0|^{1/3}\approx 2.75\; \Tch.
\ee
Thus, for the value of the chiral condensate $\braket{\bar{\psi}\psi}_0=-(250~\mbox{MeV})^3$, it predicts
\be
\Tch\approx 90~\mbox{MeV}
\ee
and this way provides a decent estimate for the chiral restoration temperature
(on the lattice this temperature in the chiral limit is around 130 MeV \cite{Karsch}), given a simplified form of the interquark interaction employed in the calculation.

\section{Conclusions}
 
In this paper we have demonstrated within a solvable chirally symmetric confining model that dynamical breaking of chiral symmetry in the vacuum at $T=0$, induced by a confining potential, disappears at some critical temperature. Then the system turns from the Nambu-Goldstone mode to the Wigner-Weyl mode featuring chirally
symmetric hadrons, confinement, and approximate chiral spin symmetry. The physical mechanism of chiral symmetry restoration is the same as in the Nambu--Jona-Lasinio model: thermal excitations of quarks and antiquarks result in Pauli blocking of the levels required for the excitation of a nonvanishing quark condensate.

\section*{Acknowledgements}

Work of A.N. was supported by the Slovenian Research Agency (research core Funding No. P1-0035) and the CAS President’s International Fellowship Initiative (PIFI) (Grant No. 2024PVA0004).

\end{document}